\documentclass[prl,twocolumn,aps,showpacs,superscriptaddress]{revtex4}

\usepackage{graphicx}  
\usepackage{dcolumn}   
\usepackage{bm}        
\usepackage{amssymb}   
\usepackage{bbold}

\usepackage{amsmath}
\usepackage{braket}
\usepackage{color}

\begin{document}

\title{Normal stresses, contraction, and stiffening in sheared elastic networks}

\author{Karsten Baumgarten}
\affiliation{Delft University of Technology, Process \& Energy Laboratory, Leeghwaterstraat 39, 2628 CB Delft, The Netherlands}

\author{Brian P. Tighe}
\affiliation{Delft University of Technology, Process \& Energy Laboratory, Leeghwaterstraat 39, 2628 CB Delft, The Netherlands}

\date{\today}

\begin{abstract}
When elastic solids are sheared, a nonlinear effect named after Poynting gives rise to normal stresses or changes in volume.
We provide a novel relation between the Poynting effect and the microscopic Gr\"uneisen parameter, which quantifies how stretching shifts vibrational modes.
By applying this relation to random spring networks, a minimal model for, e.g., biopolymer gels and solid foams, we find that networks contract  or develop tension because they vibrate faster when stretched. The amplitude of the Poynting effect is sensitive to the network's linear elastic moduli, which can be tuned via its  preparation protocol and connectivity. Finally, we show that the Poynting effect can be used to predict the finite strain scale where the material stiffens under shear.
\end{abstract}

\maketitle

\begin{figure}[b]
\centering
\includegraphics[width = 1.0\columnwidth]{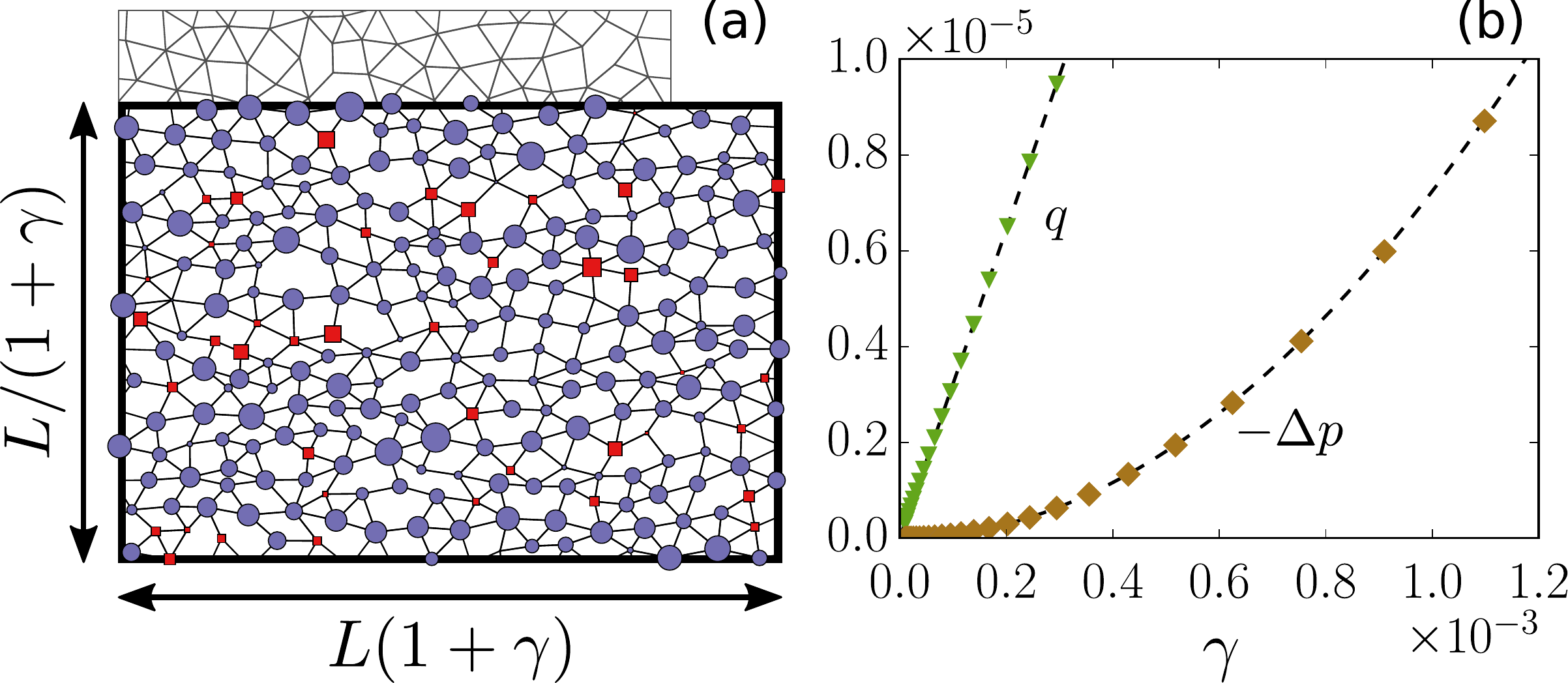}
\caption{(a) Pure shear strain $\gamma$ applied to an unstressed  spring network. Each node's area is proportional to its contribution to the pressure $p$; circles (squares) are tensile (compressive).  (b) The initial growth of the shear stress $q$ is linear in $\gamma$, while $p$ is quadratic and negative (tensile). 
}
\label{fig:intro}
\end{figure}

The Poynting effect refers to the tendency of a sheared elastic solid to expand or contract in the direction normal to a shearing surface, or to develop normal stress  if the surface is held fixed \cite{kelvin, poynting}. A similar phenomenon known as Reynolds dilatancy occurs during plastic (versus elastic) deformation of granular media \cite{reynolds1885,ren13}. The ``positive'' Poynting effect was first observed in metal wires, which lengthen or push outwards at their ends when twisted \cite{poynting}. More recently, the negative Poynting effect (contraction or tension)  was seen in  semiflexible polymers from the cytoskeleton and extracellular matrix \cite{janmey07}. 

Models of the  Poynting effect contain phenomenological elements or strong approximations \cite{weaire03,janmey07,cioroianu13,tighe14,meng16,cagny16,horgan17}.  Suggested causes in fiber networks include asymmetry (hence nonlinearity) in the fibers' force extension curve \cite{janmey07,cioroianu13}, fiber alignment in the initial condition \cite{horgan17}, and non-affinity in networks stabilized by bending \cite{heussinger07,conti09}. While these ingredients may be sufficient to induce the Poynting effect, we find that they are not necessary. This point is made by Fig.~\ref{fig:intro}a, which depicts an isotropic spring network in 2D subjected to pure shear at constant volumetric strain $\epsilon = 0$. The springs are purely harmonic and initially isotropic, and there are no bending interactions. Nodes that develop tension, labeled with a circle, greatly outnumber nodes under compression (squares), suggesting a negative Poynting effect. And indeed a plot of the pressure change $\Delta p$ is negative (Fig.~\ref{fig:intro}b). While the shear stress $q$ grows linearly with the shear strain $\gamma$, $\Delta p$ grows quadratically due to isotropy, which requires pressure or volume changes to be even in $\gamma$.

In this Letter we  introduce a new micromechanical approach to the Poynting effect, applicable for any elastic interaction between network elements. 
We focus on the initial growth of $\Delta p$ and $\epsilon$ through the coefficients
\begin{equation}
\chi_\epsilon = \left[\left(
\frac{\partial^2 p}{\partial \gamma^2}
\right)_\epsilon \right]_0
\,\,\,\,{\rm and}\,\,\,\,
\chi_p = \left[\left(
\frac{\partial^2 \epsilon}{\partial \gamma^2}
\right)_p \right]_0 \,.
\end{equation}
$\chi_\epsilon$ and $\chi_p$ are evaluated in the initial condition (``$0$"). Their subscript distinguishes strain control (fixed $\epsilon = 0$) from stress control (fixed $\Delta p = p - p_0 = 0$). We derive exact expressions for the  coefficients in hyperelastic solids (e.g.~rubbers, solid foams, and tissue), which have reversible stress-strain relations. Note that particulate media are generally not hyperelastic due to shear-induced rearrangements.
We relate $\chi_\epsilon$ and $\chi_p$ to a network's vibrational modes and the microscopic Gr\"uneisen parameter $\Gamma_n$ \cite{xu10}, which quantifies how volumetric strain shifts the frequency $\omega_n$ of the $n^{\rm th}$ mode,
\begin{equation}
\Gamma_n = - \left[\left(\frac{1}{\omega_n}
\frac{\partial \omega_n}{\partial {\epsilon}} 
\right)_\gamma \right]_0 \,.
\end{equation} 
We validate our predictions numerically in random networks of linear springs (Fig.~1), which are widely studied as minimal models of, e.g., polymer networks, foams, and glasses \cite{feng85,jacobs96,wyart08,tighe08a,ellenbroek09b,tighe12,sheinman12,lerner14,buss16}. We show that the sign of the Poynting effect in spring networks is negative and set by the  Gr\"uneisen parameter, which can be motivated theoretically.
We focus on marginally rigid spring networks close to the isostatic state (mean coordination $z = z_c + \Delta z$, with $z_c \approx 4$ in 2D), and study scaling with $\Delta z$. 

\begin{figure}
\centering
\includegraphics[width = 1.0\columnwidth]{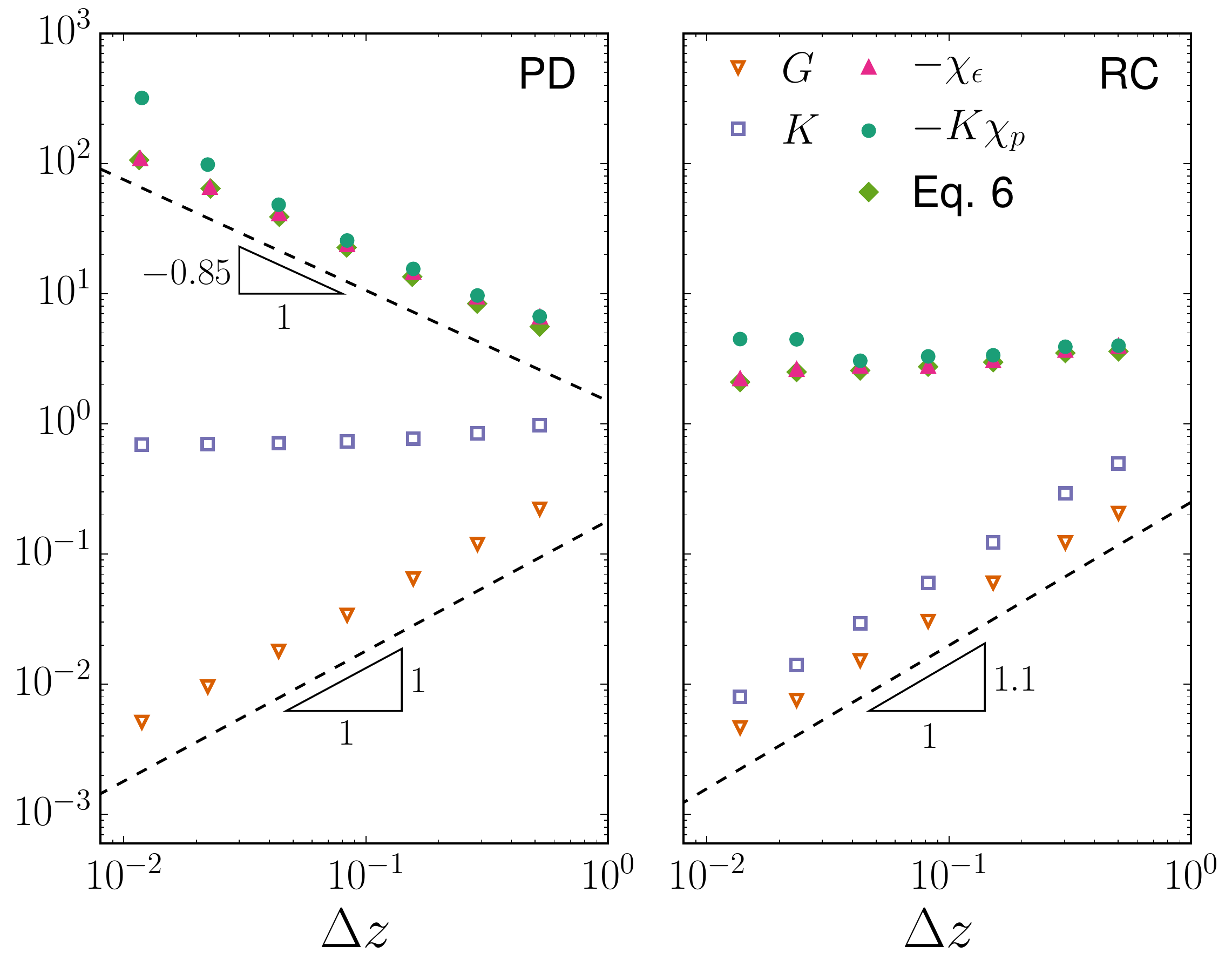}
\caption{The shear modulus $G$, bulk modulus $K$, and Poynting coefficients $\chi_\epsilon$ and $\chi_p$ as a function of excess coordination $\Delta z$ for (a) packing derived and (b) randomly cut spring networks. }
\label{fig:scaling_panel}
\end{figure}

\textit{Spring networks.---} For concreteness, we first illustrate the Poynting effect in random spring networks. 

We consider networks of $N = 1024$ harmonic springs in a periodic unit cell with initial side lengths $L_1 = L_2 = L$. Networks are prepared in two ways. Packing derived (PD) networks are prepared by  generating bidisperse packings of soft repulsive disks close to the jamming transition \cite{ohern04, vanhecke10, koeze16}. Each contact between disks is then replaced by a spring with stiffness $k$ and a rest length $\ell_{ij}^0$ equal to its initial length $\ell_{ij}$, so $p_0 = 0$ \cite{xu10,mizuno16b,buss16,baumgarten17,baumgarten17b}. To prepare randomly cut (RC) networks, we start from a PD network with mean coordination $z \approx 6$ and randomly remove springs, with a bias towards highly connected nodes \cite{wyart08,ellenbroek09b,tighe12}. All numerical results are presented in dimensionless units by setting $k$ and the average disk size in the initial packing to unity.
PD and RC networks are indistinguishable by eye, but their shear modulus $G \sim \Delta z^\mu$ and bulk modulus $K \sim \Delta z^{\mu'}$ have qualitatively different dependence on excess connectivity $\Delta z$ (Fig.~\ref{fig:scaling_panel}, open symbols). In PD networks, $G$ vanishes continuously with $\mu_{\rm PD} = 1$, while $K $ jumps discontinuously to zero ($\mu'_{\rm PD} = 0$) \cite{ohern03}. In contrast, in RC networks both $G$ and $K$ vanish continuously, with $\mu_{\rm RC} = \mu'_{\rm RC} \approx 1.1$ \cite{wyart08,ellenbroek09b,tighe12}. We will exploit these differences to test our predictions for the Poynting coefficients.

We consider deformations combining pure shear strain and volumetric expansion, such that lattice vectors of the unit cell are transformed by the deformation gradient
\begin{equation}
{\bf F} =
(1 + \epsilon )
 \left[ \begin{array}{ccc}
1 + \gamma  & 0 \\
0 & (1 + \gamma)^{-1}
\end{array}
\right] \,.
\end{equation}
The corresponding Cauchy stress tensor is 
\begin{equation}
{\bf \sigma} = \left[ \begin{array}{cc}
-p-q   & 0 \\
0 & -p+q 
\end{array}
\right] \,.
\end{equation}  
When networks are sheared using strain control, $\epsilon$ is held fixed at zero while  $\gamma$ is increased incrementally. At each step the elastic energy $\sum_{ij} V_{ij} =  (k/2) \sum_{ij} (\ell_{ij} - \ell_{ij}^0)^2$ is minimized with respect to the node positions using FIRE \cite{fire}. 
The resulting $p$ and $q$ 
are determined from
$\sigma_{\alpha\beta} = 1/(L_1 L_2)\sum_{ij} f_{ij} \, \ell_{ij} (\hat n_{ij,\alpha} \hat n_{ij,\beta})$, where $ f_{ij} = -\partial V_{ij}/\partial \ell_{ij}$ and $\hat n_{ij}$ is a unit vector pointing from node $i$ to $j$. For stress controlled simulations, $\gamma$ and $\epsilon$ are also allowed to vary while the energy is minimized subject to $p = 0$ and a prescribed $q$ \cite{dagois-bohy12}. Because finite-sized systems are never perfectly isotropic, plots of $p$ or $\epsilon$ versus $\gamma$ contain a linear contribution with a prefactor that vanishes as $N \rightarrow \infty$ \cite{goodrich14}. To  estimate the Poynting coefficients, we symmetrize $p$ and $\epsilon$ by averaging the response to shearing both ``forward'' ($\gamma > 0$) and ``backward'' ($\gamma < 0$).

Fig.~\ref{fig:scaling_panel}  presents our first main result, the Poynting coefficients for PD and RC networks over a range of $\Delta z$.  In all cases the Poynting effect is negative. There is  an apparent equality between $\chi_\epsilon$ and $K \chi_p$ (motivated  below), albeit with fluctuations at the lowest $z$. There is a notable difference in how the PD and RC Poynting coefficients scale with with $\Delta z$. In PD networks $\chi_\epsilon $ and $ K \chi_p$ diverge, with an empirical fit to $1/\Delta z^{\lambda_{\rm PD}}$ giving $\lambda_{\rm PD} \approx 0.85$.
In contrast, in  RC networks $\chi_\epsilon$ and $K \chi_p$ are  flat ($\lambda_{\rm RC} = 0$). Hence the Poynting coefficients depend on both  preparation and shearing protocols, and in three out of four cases they diverge at the isostatic point.

\textit{Microscopic theory.---} We now develop exact expressions for the Poynting coefficients, beginning with the relation between $\chi_\epsilon$ and $\chi_p$. In a hyperelastic material, the  pressure $\Delta p = (1/2) \chi_\epsilon \gamma^2$ due to shearing at fixed $\epsilon$ must be equal to the pressure from a two-step process: first shearing to $\gamma$ at constant $p$, followed by a volumetric strain $\epsilon = -(1/2)\chi_p\gamma^2$ that reverses the volume change induced in the first leg. The second step changes pressure by $\Delta p = -K\epsilon = (1/2)K\chi_p \gamma^2$, and therefore $\chi_\epsilon = K \chi_p$.

We next relate $\chi_\epsilon$ to the shear modulus $G(\epsilon) = (1/2) [(\partial q/\partial \gamma)_\epsilon]_{\gamma = 0}$ after a volumetric strain. 
The total differential of the strain energy density is ${\rm d}W = {\bf S}\!:\!{\rm d}{\bf E}$, where ${\bf E} = ({\bf F}^T{\bf F} - \mathbb{1})/2$ is the Green-Lagrange strain. The second Piola-Kirchoff stress ${\bf S}$ is related to the more experimentally-relevant Cauchy stress via ${\bf \sigma} = { {\bf F} {\bf S} {\bf F}^T}/J$, where $J = {\rm det}\,{\bf F}$. 
Hence 
\begin{equation}
{\rm d}W  = 2(1+\epsilon)^2 \left[  -\frac{p \,{\rm d}\epsilon}{1+\epsilon} + \frac{q \, {\rm d}\gamma}{1 + \gamma} \right]\,.
\label{eqn:dW}
\end{equation}
Using the Maxwell relation of Eq.~(\ref{eqn:dW}), one finds
\begin{equation}
\label{eqn:chi_e}
 \chi_\epsilon = -2G'(0) - 4G(0),
\end{equation}
where the prime indicates differentiation with respect to $\epsilon$. Earlier work neglected the difference between the various stress and strain measures in nonlinear elasticity, but still arrived at the same result \cite{weaire03,tighe14}. Numerical evaluation of Eq.~(\ref{eqn:chi_e})  is in good agreement with direct measurements of $\chi_\epsilon$ and $K \chi_p$, as shown in Fig.~\ref{fig:scaling_panel}.

We now relate $\chi_\epsilon$ to discrete degrees of freedom. Network elasticity is encoded in the the extended Hessian  ${\bf H} = \partial^2 U/\partial {\bf q}^2$, where the   the $2N+1$-component vector $\bf q$ contains the node positions and shear strain $\gamma$ \cite{tighe11}. The shear modulus can be written as a sum over the non-rigid body eigenmodes of ${\bf H}$, $1/G = (v/N) \sum_n \Lambda_n^2 / \omega_n^2$, where $v = J L^2/N$, $\omega_n^2$ is the squared eigenfrequency of the $n^{\rm th}$ eigenvector, and $\Lambda_n / N$ is its component along the strain coordinate \cite{tighe11}.
Letting $D(\omega)$, $\Lambda^2(\omega)$ and $\Gamma(\omega)$ denote the density of states and averages of  $\Lambda_n^2$ and the Gr\"uneisen parameter $\Gamma_n$ in the interval $[\omega, \omega+{\rm d}\omega)$, and replacing  sums with integrals, we find
\begin{equation}
\frac{1}{G} = v\int_0^\infty \frac{D\Lambda^2}{\omega^2} \, {\rm d}\omega \,,
\label{eqn:G}
\end{equation}
and, from Eq.~(\ref{eqn:chi_e}),
\begin{equation}
 \chi_\epsilon = 2v G^2 \int_0^\infty \
  \frac{\Gamma}{\omega^2}  \left[ 2 -  \frac{\partial \ln{\Lambda^2 D}}{\partial \ln{ \omega}} \right] 
 { \Lambda^2 D}  \, {\rm d}\omega\,.
\label{eqn:gruneisen}
\end{equation}
Eq.~(\ref{eqn:gruneisen}) is a central result: it explicitly relates the Poynting effect to vibrational modes. 
Note that the sign of $\chi_\epsilon$ is controlled by $\Gamma$ and the logarithmic derivative of $\Lambda^2 D$.

\begin{figure}[tb]
\centering
\includegraphics[width = 1.0\columnwidth]{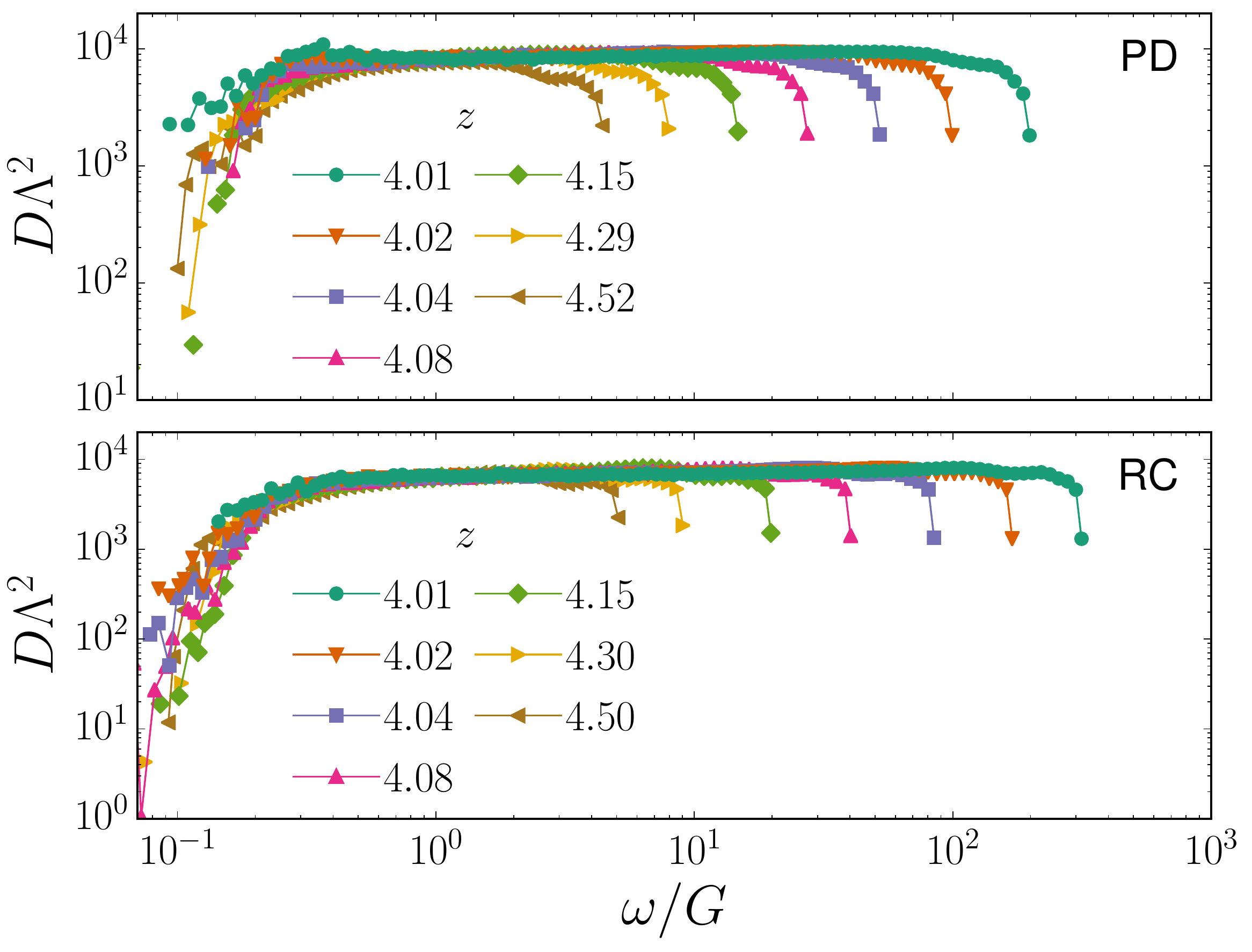}
\caption{The product $D\Lambda^2$ versus eigenfrequency $\omega$ in PD and RC networks at varying coordination $z$. $D$ is the density of states and $\Lambda^2$ is a measure of modes' coupling to shear.
}
\label{fig:overlap}
\end{figure}

\begin{figure}[tb]
\centering
\includegraphics[width = 1.0\columnwidth]{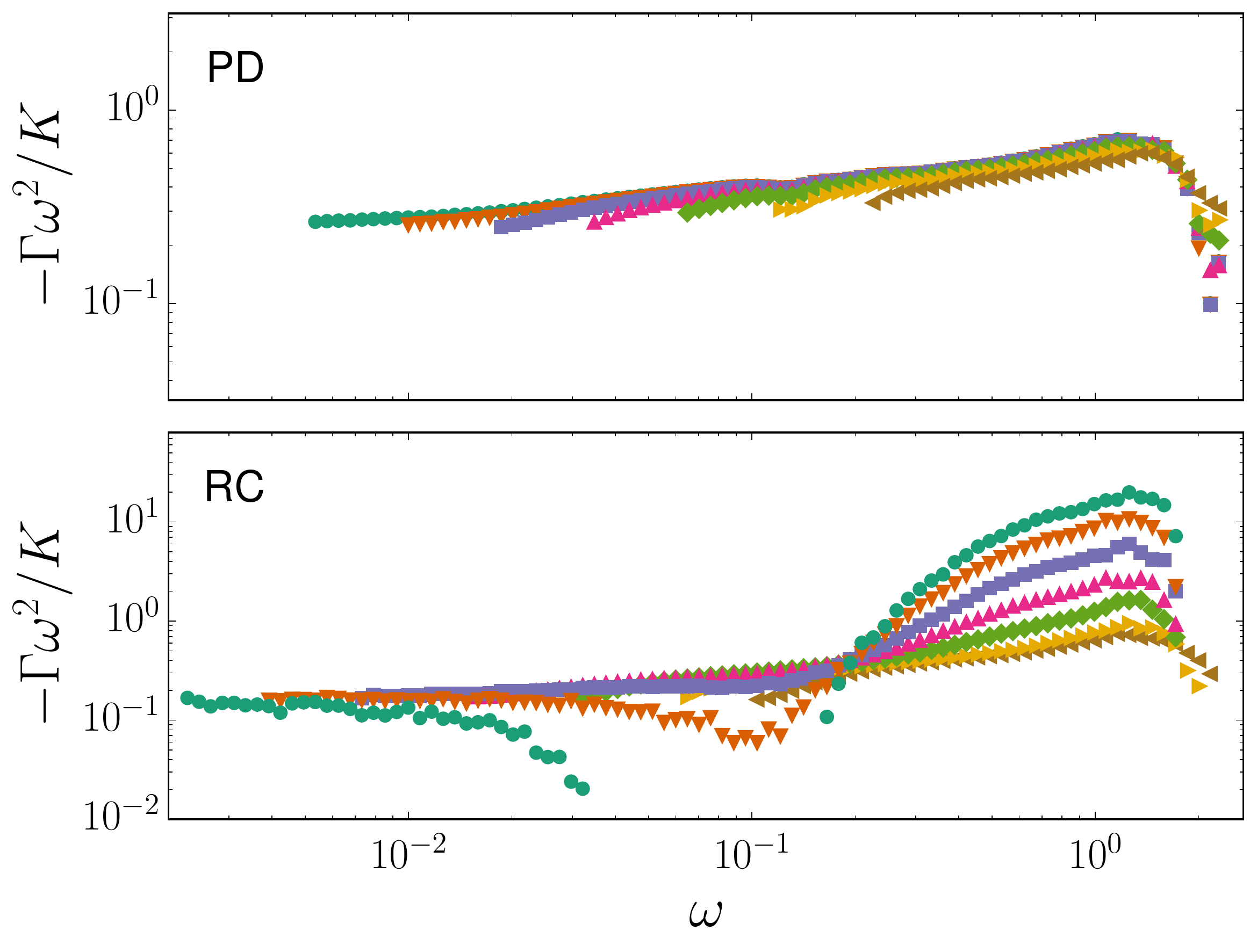}
\caption{The Gr\"uneisen parameter $\Gamma$ for PD and RD networks, scaled by the prediction of Eq.~(\ref{eqn:Gamma}) and plotted for $\omega > \omega^*$. Symbols match the legends in Fig.~\ref{fig:overlap}}
\label{fig:gruneisen}
\end{figure}

{\em Application to networks.---} We now evaluate Eq.~(\ref{eqn:gruneisen}) in the context  of spring networks, focusing on the scaling of $\chi_\epsilon$ with $\Delta z$.
Close to the isostatic state, both PD and RC networks display an anomalous abundance of ``soft modes'' that dominate the response to forcing \cite{silbert05, wyart05b, wyart08, tighe11}. The modes appear above a characteristic frequency $\omega^*$, and for scaling analysis the density of states is well approximated by a window function between $\omega^*$ and  $\omega_0 \sim {\cal O}(k^{1/2})$ \cite{silbert05, wyart05b, yan16}. Following Ref.~\cite{tighe11}, we assume that all soft modes couple similarly to shear, so  $\Lambda^2 \sim {\rm const}$. 
Hence Eqs.~(\ref{eqn:G}) and (\ref{eqn:gruneisen}) give $\omega^* \sim G$  and
\begin{equation}
\chi_\epsilon \sim G^2 \int_{\omega^*}^{\omega_0} \frac{\Gamma}{\omega^2} {\rm d}\omega \,.
\label{eqn:chiscaling}
\end{equation}

The sign and form of $\Gamma$ can be rationalized with scaling arguments. Perturbing a network along mode $n$ carries an energetic cost $\Delta U \propto \omega_n^2$, so $\Gamma \sim -\omega^{-2}(\partial \, \Delta U/\partial \epsilon)$. $\Delta U$ can be expanded in $u^\parallel_{ij}$ and $u^\perp_{ij}$, the relative normal and transverse motions,  respectively, between connected nodes. The well-known result is 
$\Delta U = (1/2)\sum_{ij} [k (u^\parallel_{ij})^2 - ({f_{ij}}/{\ell_{ij}})(u^\perp_{ij})^2 ] $,
where the force $f_{ij}$ and length $\ell_{ij}$ are evaluated prior to the perturbation \cite{alexander}. 
In a network that has previously undergone a small volumetric strain $\epsilon$ from its unstressed state, the typical force will be proportional to the pressure $p = -K\epsilon$, and so $\partial \, \Delta U/\partial \epsilon \sim K (u^\perp)^2 N$.
Soft modes strongly resemble floppy motions (which neither stretch nor compress springs), with typical transverse motions $u^\perp \sim 1/N^{1/2}$, independent of $\omega$  \cite{wyart05b, mizuno16b}.
Therefore 
\begin{equation}
\Gamma \sim  -K/\omega^2 \,,
\label{eqn:Gamma}
\end{equation}
and, by Eq.~(\ref{eqn:chiscaling}),
\begin{equation}
\chi_\epsilon \sim -K/G \,.
\end{equation}
This remarkably simple expression for $\chi_\epsilon$  correctly predicts the sign of the Poynting effect  and captures all of the phenomenology in Fig.~\ref{fig:scaling_panel}. It relates the qualitatively different behavior of $\chi_\epsilon$  in PD and RC networks  to  the differences in their shear and bulk moduli, predicting $\lambda_{\rm RC} = 0$ and $\lambda_{\rm PD} = \mu_{\rm PD} =  1$. On a qualitative level, it  explains that the Poynting effect in spring networks is negative because tension is stabilizing.
Finally, the strength of the Poynting effect grows near isostaticity because tension couples to transverse motions, which dominate soft modes and cause strong non-affine fluctuations \cite{wyart05b, wyart08, baumgarten17b}.  

The above scaling arguments rely on two essential approximations, namely that $D\Lambda^2 \sim {\rm const}$ and $\Gamma \sim -K/\omega^2$ above $\omega^* \sim G$.
We now validate them by direct numerical evaluation.  In Fig.~\ref{fig:overlap}, $D\Lambda^2$ is plotted as a function of $\omega/G$ for both PD and RC networks. As expected, in both cases there is  a broad plateau above $\omega^*$. In Fig.~\ref{fig:gruneisen} we plot the ratio of  $\Gamma$ to $-K/\omega^2$; $\Gamma$ is estimated from a linear fit of $\omega_n$ versus $\epsilon$ after a series of small volumetric strain steps. In PD networks the ratio approaches a positive constant as $\Delta z \rightarrow 0$, indicating that Eq.~(\ref{eqn:Gamma}) becomes increasingly accurate as the isostatic point is approached.  At finite $\Delta z$ there is a slow upturn with increasing $\omega$. We attribute this to a subdominant correction to scaling, consistent with the observation that a power law fit to $\chi_\epsilon$ and $K \chi_p$ in PD networks gives a somewhat smaller value of $\lambda_{\rm PD}$ than 1. The same ratio has a more complex form in RC networks, including a sign change for the lowest $z$, but it also approaches a low frequency plateau in the isostatic limit.

\begin{figure}
 \centering
 \includegraphics[width = 1.0\columnwidth]{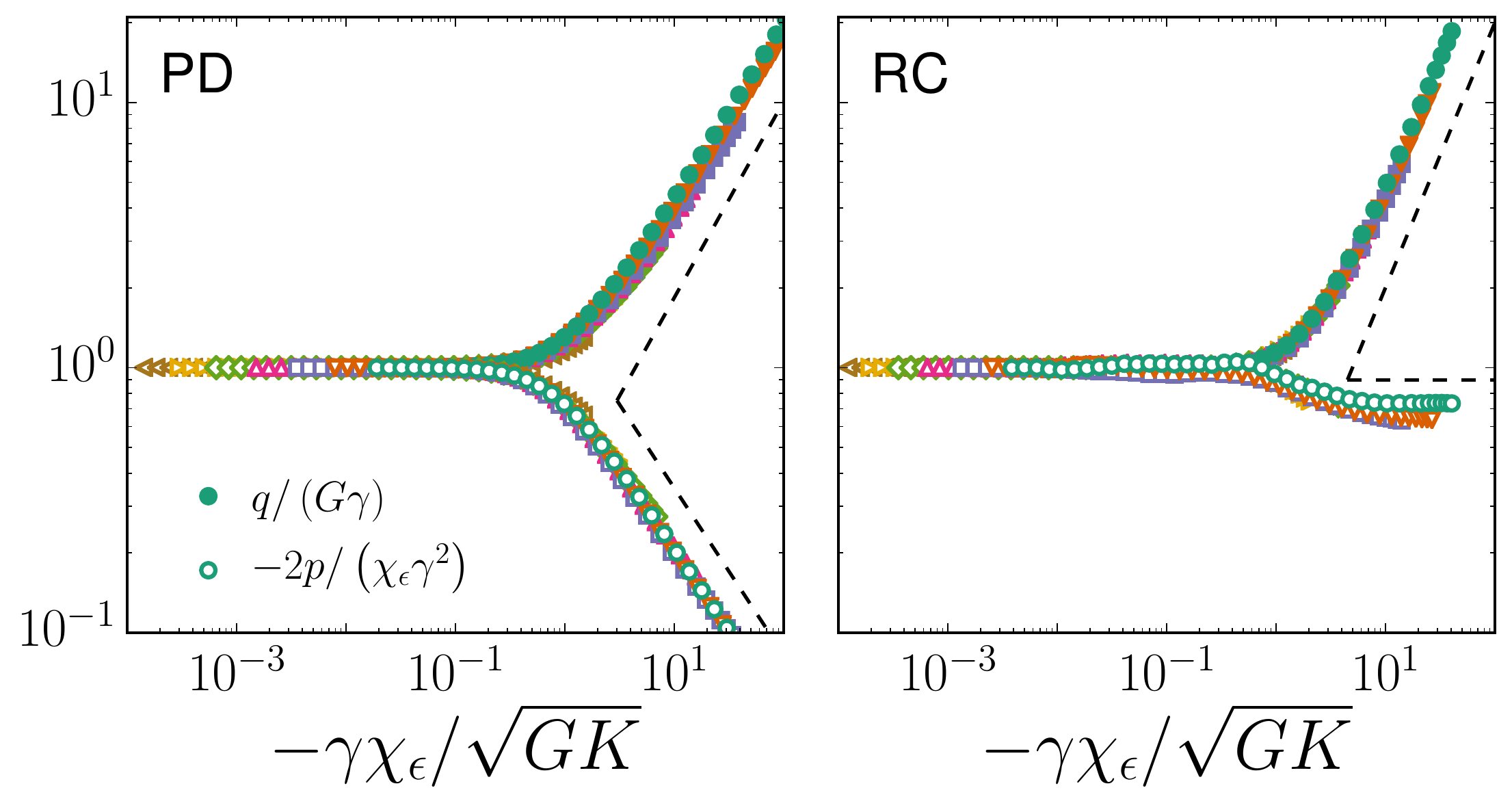}
 \caption{Master curves for shear stress $q$ and pressure $p$ of PD and RC networks sheared to finite strain $\gamma$ at fixed $\epsilon = 0$. The dashed lines on the left have slopes $0.74$ and $-0.63$. On the right the slopes are one and zero.}
 \label{fig:stiffening}
\end{figure}
\textit{Finite strain.---} The Poynting coefficients quantify the leading order dependence of $\Delta p$ and $\epsilon$ on $\gamma$. We now show that the Poynting coefficient $\chi_{\epsilon}$ predicts the onset of strain stiffening when a network is sheared at fixed volume.

There has been no prior study of PD networks at finite strain, while studies of RC networks did not report normal stresses. RC shear stresses were  shown to stiffen beyond some vanishing strain scale $\gamma^*$ \cite{wyart08} (unlike sphere packings, which soften \cite{boschan16,dagoisbohy17}). The secant modulus $q/\gamma$ in RC networks satisfies $q/(G\gamma) = {\cal Q}(\gamma/\gamma^*)$, with ${\cal Q} \sim 1$ for $x \ll 1$ and ${\cal Q} \sim |x|^{\theta}$ with $\theta > 0$ for $x \gg 1$ \cite{wyart08}.
 It is natural to make a similar ansatz for the pressure,
 \begin{equation}
\frac{2p}{\chi_\epsilon \gamma^2} = {\cal P}(\gamma/\gamma^*) \,,
 \end{equation}
 where ${\cal P} \sim 1$ for $x \ll 1$ and ${\cal P} \sim |x|^{\phi}$ for $x \gg 1$. 

The scaling functions ${\cal Q}$ and ${\cal P}$  are plotted in Fig.~\ref{fig:stiffening}. In Ref.~\cite{wyart08} it was argued that $\gamma^* \sim \Delta z$, which agrees with our  RC network data but fails for PD networks. Instead, we find that data from both network types collapses with

\begin{equation}\label{eqn:gammastar}
\gamma^* \sim \frac{\sqrt{GK}}{|\chi_\epsilon|} \sim 
\Delta z^{\nu} \,,
\end{equation}
with $\nu = \lambda + (\mu + \mu')/2$.
In order for shear stress and pressure to remain finite when $\Delta z \rightarrow 0$, we must have 
$\theta = \mu/\nu$ and $\phi = -\lambda/\nu$. 
These give $\theta_{\rm PD} \approx 0.74$ and $\phi_{\rm PD} \approx -0.63$  (using $\lambda_{\rm PD} = 0.85$), as well as $\theta_{\rm RC} \approx 1$ and $\phi_{\rm RC} \approx 0$. These are all in good agreement with numerics (dashed lines in Fig.~\ref{fig:stiffening}).
  
To motivate $\gamma^*$, we expand the secant modulus in $p(\gamma)$ to find
 \begin{equation}
 \frac{q}{2G\gamma} \ \sim 1 + \frac{\chi_{\epsilon}^2\gamma^2}{GK} + \mathcal{O}\left(\gamma^4\right)\,.
 \label{eqn:expansion}
 \end{equation}
Here we have neglected numerical prefactors and used Eq.~(\ref{eqn:chi_e}), assuming $G'(0) \gg G(0)$ (appropriate near isostaticity). Balancing terms on the right-hand side yields Eq.~(\ref{eqn:gammastar}), an extrapolated strain scale where  the initial linear form of the stress-strain curve breaks down. A link between normal stresses and stiffening was also evidenced in \cite{vahabi17}.

\textit{Conclusion--}
We have derived exact expressions for the Poynting coefficients in hyperelastic materials, and validated them numerically in two classes of spring networks.  Both display a negative Poynting effect, whose origin can be traced to the stabilizing influence of tension on a network's vibrational modes. The amplitude of the effect is controlled by the coupling between tension and relative transverse motions, which explains the correlation between normal stress and non-affinity \cite{conti09}, and results macroscopically in a coefficient $\chi_\epsilon$ that scales with the ratio $K/G$. 
Eq.~(\ref{eqn:gruneisen}) is applicable in any 2D hyperelastic material -- hence our results can lend insight to the Poynting effect in other elastic networks, including fiber networks (e.g.~\cite{astrom00,wilhelm03,head03,heussinger07,conti09,das12}). The scaling arguments for $D\Lambda^2$ and $\Gamma$ presented here are specific to spring networks; they must be modeled or evaluated anew for each material.
Our calculations and numerics are all in 2D, but extension to 3D is straightforward and we do not expect the underlying physics to change. 

We have shown that  Poynting coefficients and stiffening behavior are highly sensitive to the linear elastic moduli. Recent work has demonstrated how to prepare spring networks using a biased cutting protocol to target  essentially any positive value of $K/G$ \cite{goodrich15,reid18}. Our results indicate that the same techniques could be used to select for desirable nonlinear mechanical properties.
 
One can ask whether the elastic Poynting effect gives insight into Reynolds dilatancy. As noted above, our approach does not apply to irreversible deformations. More heuristically, we note that whereas volumetric expansion stabilizes elastic systems, it destabilizes particulate matter by opening contacts. This suggests a sign difference, and indeed materials like sand generally dilate under shear unless prepared in a loose state.

{\em Acknowledgments.---} 
We acknowledge financial support from the Netherlands Organization for Scientific Research (NWO) and the use of supercomputer facilities provided by NWO Physical Sciences.

\bibliographystyle{apsrev}

\begin{thebibliography}{47}
\expandafter\ifx\csname natexlab\endcsname\relax\def\natexlab#1{#1}\fi
\expandafter\ifx\csname bibnamefont\endcsname\relax
  \def\bibnamefont#1{#1}\fi
\expandafter\ifx\csname bibfnamefont\endcsname\relax
  \def\bibfnamefont#1{#1}\fi
\expandafter\ifx\csname citenamefont\endcsname\relax
  \def\citenamefont#1{#1}\fi
\expandafter\ifx\csname url\endcsname\relax
  \def\url#1{\texttt{#1}}\fi
\expandafter\ifx\csname urlprefix\endcsname\relax\def\urlprefix{URL }\fi
\providecommand{\bibinfo}[2]{#2}
\providecommand{\eprint}[2][]{\url{#2}}

\bibitem[{\citenamefont{Thomson}(1878)}]{kelvin}
\bibinfo{author}{\bibfnamefont{W.}~\bibnamefont{Thomson}}, in
  \emph{\bibinfo{booktitle}{Encyclopaedia Britannica}}
  (\bibinfo{publisher}{Adam and Charles Black, Edinburgh},
  \bibinfo{year}{1878}).

\bibitem[{\citenamefont{Poynting}(1909)}]{poynting}
\bibinfo{author}{\bibfnamefont{J.}~\bibnamefont{Poynting}},
  \bibinfo{journal}{Proceedings of the Royal Society of London}
  \textbf{\bibinfo{volume}{82}}, \bibinfo{pages}{546} (\bibinfo{year}{1909}).

\bibitem[{\citenamefont{Reynolds}(1885)}]{reynolds1885}
\bibinfo{author}{\bibfnamefont{O.}~\bibnamefont{Reynolds}}, in
  \emph{\bibinfo{booktitle}{Proc. Brit. Assoc.}} (\bibinfo{year}{1885}), p.
  \bibinfo{pages}{896}.

\bibitem[{\citenamefont{Ren et~al.}(2013)\citenamefont{Ren, Dijksman, and
  Behringer}}]{ren13}
\bibinfo{author}{\bibfnamefont{J.}~\bibnamefont{Ren}},
  \bibinfo{author}{\bibfnamefont{J.~A.} \bibnamefont{Dijksman}},
  \bibnamefont{and} \bibinfo{author}{\bibfnamefont{R.~P.}
  \bibnamefont{Behringer}}, \bibinfo{journal}{Phys. Rev. Lett.}
  \textbf{\bibinfo{volume}{110}}, \bibinfo{pages}{018302}
  (\bibinfo{year}{2013}).

\bibitem[{\citenamefont{Janmey et~al.}(2007)\citenamefont{Janmey, McCormick,
  Rammensee, Leight, Georges, and MacKintosh}}]{janmey07}
\bibinfo{author}{\bibfnamefont{P.~A.} \bibnamefont{Janmey}},
  \bibinfo{author}{\bibfnamefont{M.~E.} \bibnamefont{McCormick}},
  \bibinfo{author}{\bibfnamefont{S.}~\bibnamefont{Rammensee}},
  \bibinfo{author}{\bibfnamefont{J.~L.} \bibnamefont{Leight}},
  \bibinfo{author}{\bibfnamefont{P.~C.} \bibnamefont{Georges}},
  \bibnamefont{and} \bibinfo{author}{\bibfnamefont{F.~C.}
  \bibnamefont{MacKintosh}}, \bibinfo{journal}{Nature Materials}
  \textbf{\bibinfo{volume}{6}}, \bibinfo{pages}{48 } (\bibinfo{year}{2007}).

\bibitem[{\citenamefont{Weaire and Hutzler}(2003)}]{weaire03}
\bibinfo{author}{\bibfnamefont{D.}~\bibnamefont{Weaire}} \bibnamefont{and}
  \bibinfo{author}{\bibfnamefont{S.}~\bibnamefont{Hutzler}},
  \bibinfo{journal}{Philosoph.~Mag.} \textbf{\bibinfo{volume}{83}},
  \bibinfo{pages}{2747} (\bibinfo{year}{2003}).

\bibitem[{\citenamefont{Cioroianu and Storm}(2013)}]{cioroianu13}
\bibinfo{author}{\bibfnamefont{A.~R.} \bibnamefont{Cioroianu}}
  \bibnamefont{and} \bibinfo{author}{\bibfnamefont{C.}~\bibnamefont{Storm}},
  \bibinfo{journal}{Phys. Rev. E} \textbf{\bibinfo{volume}{88}},
  \bibinfo{pages}{052601} (\bibinfo{year}{2013}).

\bibitem[{\citenamefont{Tighe}(2014)}]{tighe14}
\bibinfo{author}{\bibfnamefont{B.~P.} \bibnamefont{Tighe}},
  \bibinfo{journal}{Granular Matter} \textbf{\bibinfo{volume}{16}},
  \bibinfo{pages}{203} (\bibinfo{year}{2014}).

\bibitem[{\citenamefont{Meng and Terentjev}(2016)}]{meng16}
\bibinfo{author}{\bibfnamefont{F.}~\bibnamefont{Meng}} \bibnamefont{and}
  \bibinfo{author}{\bibfnamefont{E.~M.} \bibnamefont{Terentjev}},
  \bibinfo{journal}{Soft Matter} \textbf{\bibinfo{volume}{12}},
  \bibinfo{pages}{6749} (\bibinfo{year}{2016}).

\bibitem[{\citenamefont{de~Cagny et~al.}(2016)\citenamefont{de~Cagny, Vos,
  Vahabi, Kurniawan, Doi, Koenderink, MacKintosh, and Bonn}}]{cagny16}
\bibinfo{author}{\bibfnamefont{H.~C.} \bibnamefont{de~Cagny}},
  \bibinfo{author}{\bibfnamefont{B.~E.} \bibnamefont{Vos}},
  \bibinfo{author}{\bibfnamefont{M.}~\bibnamefont{Vahabi}},
  \bibinfo{author}{\bibfnamefont{N.~A.} \bibnamefont{Kurniawan}},
  \bibinfo{author}{\bibfnamefont{M.}~\bibnamefont{Doi}},
  \bibinfo{author}{\bibfnamefont{G.~H.} \bibnamefont{Koenderink}},
  \bibinfo{author}{\bibfnamefont{F.~C.} \bibnamefont{MacKintosh}},
  \bibnamefont{and} \bibinfo{author}{\bibfnamefont{D.}~\bibnamefont{Bonn}},
  \bibinfo{journal}{Phys. Rev. Lett.} \textbf{\bibinfo{volume}{117}},
  \bibinfo{pages}{217802} (\bibinfo{year}{2016}).

\bibitem[{\citenamefont{Horgan and Murphy}(2017)}]{horgan17}
\bibinfo{author}{\bibfnamefont{C.}~\bibnamefont{Horgan}} \bibnamefont{and}
  \bibinfo{author}{\bibfnamefont{J.}~\bibnamefont{Murphy}},
  \bibinfo{journal}{Soft Matter} \textbf{\bibinfo{volume}{13}},
  \bibinfo{pages}{4916} (\bibinfo{year}{2017}).

\bibitem[{\citenamefont{Heussinger et~al.}(2007)\citenamefont{Heussinger,
  Schaefer, and Frey}}]{heussinger07}
\bibinfo{author}{\bibfnamefont{C.}~\bibnamefont{Heussinger}},
  \bibinfo{author}{\bibfnamefont{B.}~\bibnamefont{Schaefer}}, \bibnamefont{and}
  \bibinfo{author}{\bibfnamefont{E.}~\bibnamefont{Frey}},
  \bibinfo{journal}{Phys. Rev. E} \textbf{\bibinfo{volume}{76}},
  \bibinfo{pages}{031906} (\bibinfo{year}{2007}).

\bibitem[{\citenamefont{Conti and MacKintosh}(2009)}]{conti09}
\bibinfo{author}{\bibfnamefont{E.}~\bibnamefont{Conti}} \bibnamefont{and}
  \bibinfo{author}{\bibfnamefont{F.~C.} \bibnamefont{MacKintosh}},
  \bibinfo{journal}{Phys. Rev. Lett.} \textbf{\bibinfo{volume}{102}},
  \bibinfo{pages}{088102} (\bibinfo{year}{2009}).

\bibitem[{\citenamefont{Xu et~al.}(2010)\citenamefont{Xu, Vitelli, Liu, and
  Nagel}}]{xu10}
\bibinfo{author}{\bibfnamefont{N.}~\bibnamefont{Xu}},
  \bibinfo{author}{\bibfnamefont{V.}~\bibnamefont{Vitelli}},
  \bibinfo{author}{\bibfnamefont{A.~J.} \bibnamefont{Liu}}, \bibnamefont{and}
  \bibinfo{author}{\bibfnamefont{S.~R.} \bibnamefont{Nagel}},
  \bibinfo{journal}{EPL} \textbf{\bibinfo{volume}{90}}, \bibinfo{pages}{56001}
  (\bibinfo{year}{2010}).

\bibitem[{\citenamefont{Feng et~al.}(1985)\citenamefont{Feng, Thorpe, and
  Garboczi}}]{feng85}
\bibinfo{author}{\bibfnamefont{S.}~\bibnamefont{Feng}},
  \bibinfo{author}{\bibfnamefont{M.}~\bibnamefont{Thorpe}}, \bibnamefont{and}
  \bibinfo{author}{\bibfnamefont{E.}~\bibnamefont{Garboczi}},
  \bibinfo{journal}{Physical Review B} \textbf{\bibinfo{volume}{31}},
  \bibinfo{pages}{276} (\bibinfo{year}{1985}).

\bibitem[{\citenamefont{Jacobs and Thorpe}(1996)}]{jacobs96}
\bibinfo{author}{\bibfnamefont{D.~J.} \bibnamefont{Jacobs}} \bibnamefont{and}
  \bibinfo{author}{\bibfnamefont{M.~F.} \bibnamefont{Thorpe}},
  \bibinfo{journal}{Phys. Rev. E} \textbf{\bibinfo{volume}{53}},
  \bibinfo{pages}{3682} (\bibinfo{year}{1996}).

\bibitem[{\citenamefont{Wyart et~al.}(2008)\citenamefont{Wyart, Liang, Kabla,
  and Mahadevan}}]{wyart08}
\bibinfo{author}{\bibfnamefont{M.}~\bibnamefont{Wyart}},
  \bibinfo{author}{\bibfnamefont{H.}~\bibnamefont{Liang}},
  \bibinfo{author}{\bibfnamefont{A.}~\bibnamefont{Kabla}}, \bibnamefont{and}
  \bibinfo{author}{\bibfnamefont{L.}~\bibnamefont{Mahadevan}},
  \bibinfo{journal}{Phys. Rev. Lett.} \textbf{\bibinfo{volume}{101}},
  \bibinfo{pages}{215501} (\bibinfo{year}{2008}).

\bibitem[{\citenamefont{Tighe and Socolar}(2008)}]{tighe08a}
\bibinfo{author}{\bibfnamefont{B.~P.} \bibnamefont{Tighe}} \bibnamefont{and}
  \bibinfo{author}{\bibfnamefont{J.~E.~S.} \bibnamefont{Socolar}},
  \bibinfo{journal}{Phys.~Rev.~E} \textbf{\bibinfo{volume}{77}},
  \bibinfo{pages}{031303} (\bibinfo{year}{2008}).

\bibitem[{\citenamefont{Ellenbroek et~al.}(2009)\citenamefont{Ellenbroek,
  Zeravcic, van Saarloos, and van Hecke}}]{ellenbroek09b}
\bibinfo{author}{\bibfnamefont{W.~G.} \bibnamefont{Ellenbroek}},
  \bibinfo{author}{\bibfnamefont{Z.}~\bibnamefont{Zeravcic}},
  \bibinfo{author}{\bibfnamefont{W.}~\bibnamefont{van Saarloos}},
  \bibnamefont{and} \bibinfo{author}{\bibfnamefont{M.}~\bibnamefont{van
  Hecke}}, \bibinfo{journal}{EPL} \textbf{\bibinfo{volume}{87}},
  \bibinfo{pages}{34004} (\bibinfo{year}{2009}).

\bibitem[{\citenamefont{Tighe}(2012)}]{tighe12}
\bibinfo{author}{\bibfnamefont{B.~P.} \bibnamefont{Tighe}},
  \bibinfo{journal}{Phys. Rev. Lett.} \textbf{\bibinfo{volume}{109}},
  \bibinfo{pages}{168303} (\bibinfo{year}{2012}).

\bibitem[{\citenamefont{Sheinman et~al.}(2012)\citenamefont{Sheinman,
  Broedersz, and MacKintosh}}]{sheinman12}
\bibinfo{author}{\bibfnamefont{M.}~\bibnamefont{Sheinman}},
  \bibinfo{author}{\bibfnamefont{C.~P.} \bibnamefont{Broedersz}},
  \bibnamefont{and} \bibinfo{author}{\bibfnamefont{F.~C.}
  \bibnamefont{MacKintosh}}, \bibinfo{journal}{Phys. Rev. E}
  \textbf{\bibinfo{volume}{85}}, \bibinfo{pages}{021801}
  (\bibinfo{year}{2012}).

\bibitem[{\citenamefont{Lerner et~al.}(2014)\citenamefont{Lerner, DeGiuli,
  D{\"u}ring, and Wyart}}]{lerner14}
\bibinfo{author}{\bibfnamefont{E.}~\bibnamefont{Lerner}},
  \bibinfo{author}{\bibfnamefont{E.}~\bibnamefont{DeGiuli}},
  \bibinfo{author}{\bibfnamefont{G.}~\bibnamefont{D{\"u}ring}},
  \bibnamefont{and} \bibinfo{author}{\bibfnamefont{M.}~\bibnamefont{Wyart}},
  \bibinfo{journal}{Soft Matter} \textbf{\bibinfo{volume}{10}},
  \bibinfo{pages}{5085} (\bibinfo{year}{2014}).

\bibitem[{\citenamefont{Buss et~al.}(2016)\citenamefont{Buss, Heussinger, and
  Hallatschek}}]{buss16}
\bibinfo{author}{\bibfnamefont{C.}~\bibnamefont{Buss}},
  \bibinfo{author}{\bibfnamefont{C.}~\bibnamefont{Heussinger}},
  \bibnamefont{and}
  \bibinfo{author}{\bibfnamefont{O.}~\bibnamefont{Hallatschek}},
  \bibinfo{journal}{Soft Matter} \textbf{\bibinfo{volume}{12}},
  \bibinfo{pages}{7682} (\bibinfo{year}{2016}).

\bibitem[{\citenamefont{O'Hern et~al.}(2004)\citenamefont{O'Hern, Silbert, Liu,
  and Nagel}}]{ohern04}
\bibinfo{author}{\bibfnamefont{C.~S.} \bibnamefont{O'Hern}},
  \bibinfo{author}{\bibfnamefont{L.~E.} \bibnamefont{Silbert}},
  \bibinfo{author}{\bibfnamefont{A.~J.} \bibnamefont{Liu}}, \bibnamefont{and}
  \bibinfo{author}{\bibfnamefont{S.~R.} \bibnamefont{Nagel}},
  \bibinfo{journal}{Phys. Rev. E} \textbf{\bibinfo{volume}{70}},
  \bibinfo{pages}{043302} (\bibinfo{year}{2004}).

\bibitem[{\citenamefont{van Hecke}(2010)}]{vanhecke10}
\bibinfo{author}{\bibfnamefont{M.}~\bibnamefont{van Hecke}},
  \bibinfo{journal}{J.~Phys.~Cond.~Matt.} \textbf{\bibinfo{volume}{22}},
  \bibinfo{pages}{033101} (\bibinfo{year}{2010}).

\bibitem[{\citenamefont{Koeze et~al.}(2016)\citenamefont{Koeze, V{\aa}gberg,
  Tjoa, and Tighe}}]{koeze16}
\bibinfo{author}{\bibfnamefont{D.~J.} \bibnamefont{Koeze}},
  \bibinfo{author}{\bibfnamefont{D.}~\bibnamefont{V{\aa}gberg}},
  \bibinfo{author}{\bibfnamefont{B.~B.} \bibnamefont{Tjoa}}, \bibnamefont{and}
  \bibinfo{author}{\bibfnamefont{B.~P.} \bibnamefont{Tighe}},
  \bibinfo{journal}{EPL} \textbf{\bibinfo{volume}{113}}, \bibinfo{pages}{54001}
  (\bibinfo{year}{2016}).

\bibitem[{\citenamefont{Mizuno et~al.}(2016)\citenamefont{Mizuno, Saitoh, and
  Silbert}}]{mizuno16b}
\bibinfo{author}{\bibfnamefont{H.}~\bibnamefont{Mizuno}},
  \bibinfo{author}{\bibfnamefont{K.}~\bibnamefont{Saitoh}}, \bibnamefont{and}
  \bibinfo{author}{\bibfnamefont{L.~E.} \bibnamefont{Silbert}},
  \bibinfo{journal}{Physical Review E} \textbf{\bibinfo{volume}{93}},
  \bibinfo{pages}{062905} (\bibinfo{year}{2016}).

\bibitem[{\citenamefont{Baumgarten et~al.}(2017)\citenamefont{Baumgarten,
  V{\aa}gberg, and Tighe}}]{baumgarten17}
\bibinfo{author}{\bibfnamefont{K.}~\bibnamefont{Baumgarten}},
  \bibinfo{author}{\bibfnamefont{D.}~\bibnamefont{V{\aa}gberg}},
  \bibnamefont{and} \bibinfo{author}{\bibfnamefont{B.~P.} \bibnamefont{Tighe}},
  \bibinfo{journal}{Phys. Rev. Lett.} \textbf{\bibinfo{volume}{118}},
  \bibinfo{pages}{098001} (\bibinfo{year}{2017}).

\bibitem[{\citenamefont{Baumgarten and Tighe}(2017)}]{baumgarten17b}
\bibinfo{author}{\bibfnamefont{K.}~\bibnamefont{Baumgarten}} \bibnamefont{and}
  \bibinfo{author}{\bibfnamefont{B.~P.} \bibnamefont{Tighe}},
  \bibinfo{journal}{Soft Matter} \textbf{\bibinfo{volume}{13}},
  \bibinfo{pages}{9036 } (\bibinfo{year}{2017}).

\bibitem[{\citenamefont{O'Hern et~al.}(2003)\citenamefont{O'Hern, Silbert, Liu,
  and Nagel}}]{ohern03}
\bibinfo{author}{\bibfnamefont{C.~S.} \bibnamefont{O'Hern}},
  \bibinfo{author}{\bibfnamefont{L.~E.} \bibnamefont{Silbert}},
  \bibinfo{author}{\bibfnamefont{A.~J.} \bibnamefont{Liu}}, \bibnamefont{and}
  \bibinfo{author}{\bibfnamefont{S.~R.} \bibnamefont{Nagel}},
  \bibinfo{journal}{Phys.~Rev.~E} \textbf{\bibinfo{volume}{68}},
  \bibinfo{pages}{011306} (\bibinfo{year}{2003}).

\bibitem[{\citenamefont{Bitzek et~al.}(2006)\citenamefont{Bitzek, Koskinen,
  G\"ahler, Moseler, and Gumbsch}}]{fire}
\bibinfo{author}{\bibfnamefont{E.}~\bibnamefont{Bitzek}},
  \bibinfo{author}{\bibfnamefont{P.}~\bibnamefont{Koskinen}},
  \bibinfo{author}{\bibfnamefont{F.}~\bibnamefont{G\"ahler}},
  \bibinfo{author}{\bibfnamefont{M.}~\bibnamefont{Moseler}}, \bibnamefont{and}
  \bibinfo{author}{\bibfnamefont{P.}~\bibnamefont{Gumbsch}},
  \bibinfo{journal}{Phys. Rev. Lett.} \textbf{\bibinfo{volume}{97}},
  \bibinfo{pages}{170201} (\bibinfo{year}{2006}).

\bibitem[{\citenamefont{Dagois-Bohy et~al.}(2012)\citenamefont{Dagois-Bohy,
  Tighe, Simon, Henkes, and van Hecke}}]{dagois-bohy12}
\bibinfo{author}{\bibfnamefont{S.}~\bibnamefont{Dagois-Bohy}},
  \bibinfo{author}{\bibfnamefont{B.~P.} \bibnamefont{Tighe}},
  \bibinfo{author}{\bibfnamefont{J.}~\bibnamefont{Simon}},
  \bibinfo{author}{\bibfnamefont{S.}~\bibnamefont{Henkes}}, \bibnamefont{and}
  \bibinfo{author}{\bibfnamefont{M.}~\bibnamefont{van Hecke}},
  \bibinfo{journal}{Phys. Rev. Lett.} \textbf{\bibinfo{volume}{109}},
  \bibinfo{pages}{095703} (\bibinfo{year}{2012}).

\bibitem[{\citenamefont{Goodrich et~al.}(2014)\citenamefont{Goodrich,
  Dagois-Bohy, Tighe, van Hecke, Liu, and Nagel}}]{goodrich14}
\bibinfo{author}{\bibfnamefont{C.~P.} \bibnamefont{Goodrich}},
  \bibinfo{author}{\bibfnamefont{S.}~\bibnamefont{Dagois-Bohy}},
  \bibinfo{author}{\bibfnamefont{B.~P.} \bibnamefont{Tighe}},
  \bibinfo{author}{\bibfnamefont{M.}~\bibnamefont{van Hecke}},
  \bibinfo{author}{\bibfnamefont{A.~J.} \bibnamefont{Liu}}, \bibnamefont{and}
  \bibinfo{author}{\bibfnamefont{S.~R.} \bibnamefont{Nagel}},
  \bibinfo{journal}{Phys. Rev. E} \textbf{\bibinfo{volume}{90}},
  \bibinfo{pages}{022138} (\bibinfo{year}{2014}).

\bibitem[{\citenamefont{Tighe}(2011)}]{tighe11}
\bibinfo{author}{\bibfnamefont{B.~P.} \bibnamefont{Tighe}},
  \bibinfo{journal}{Phys. Rev. Lett.} \textbf{\bibinfo{volume}{107}},
  \bibinfo{pages}{158303} (\bibinfo{year}{2011}).

\bibitem[{\citenamefont{Silbert et~al.}(2005)\citenamefont{Silbert, Liu, and
  Nagel}}]{silbert05}
\bibinfo{author}{\bibfnamefont{L.~E.} \bibnamefont{Silbert}},
  \bibinfo{author}{\bibfnamefont{A.~J.} \bibnamefont{Liu}}, \bibnamefont{and}
  \bibinfo{author}{\bibfnamefont{S.~R.} \bibnamefont{Nagel}},
  \bibinfo{journal}{Phys.~Rev.~Lett.} \textbf{\bibinfo{volume}{95}},
  \bibinfo{pages}{098301} (\bibinfo{year}{2005}).

\bibitem[{\citenamefont{Wyart et~al.}(2005)\citenamefont{Wyart, Silbert, Nagel,
  and Witten}}]{wyart05b}
\bibinfo{author}{\bibfnamefont{M.}~\bibnamefont{Wyart}},
  \bibinfo{author}{\bibfnamefont{L.~E.} \bibnamefont{Silbert}},
  \bibinfo{author}{\bibfnamefont{S.~R.} \bibnamefont{Nagel}}, \bibnamefont{and}
  \bibinfo{author}{\bibfnamefont{T.~A.} \bibnamefont{Witten}},
  \bibinfo{journal}{Phys.~Rev.~E} \textbf{\bibinfo{volume}{72}},
  \bibinfo{pages}{051306} (\bibinfo{year}{2005}).

\bibitem[{\citenamefont{Yan et~al.}(2016)\citenamefont{Yan, DeGiuli, and
  Wyart}}]{yan16}
\bibinfo{author}{\bibfnamefont{L.}~\bibnamefont{Yan}},
  \bibinfo{author}{\bibfnamefont{E.}~\bibnamefont{DeGiuli}}, \bibnamefont{and}
  \bibinfo{author}{\bibfnamefont{M.}~\bibnamefont{Wyart}},
  \bibinfo{journal}{EPL (Europhysics Letters)} \textbf{\bibinfo{volume}{114}},
  \bibinfo{pages}{26003} (\bibinfo{year}{2016}).

\bibitem[{\citenamefont{Alexander}(1998)}]{alexander}
\bibinfo{author}{\bibfnamefont{S.}~\bibnamefont{Alexander}},
  \bibinfo{journal}{Phys.~Rep} \textbf{\bibinfo{volume}{296}},
  \bibinfo{pages}{65} (\bibinfo{year}{1998}).

\bibitem[{\citenamefont{Boschan et~al.}(2016)\citenamefont{Boschan,
  V{\aa}gberg, Somfai, and Tighe}}]{boschan16}
\bibinfo{author}{\bibfnamefont{J.}~\bibnamefont{Boschan}},
  \bibinfo{author}{\bibfnamefont{D.}~\bibnamefont{V{\aa}gberg}},
  \bibinfo{author}{\bibfnamefont{E.}~\bibnamefont{Somfai}}, \bibnamefont{and}
  \bibinfo{author}{\bibfnamefont{B.~P.} \bibnamefont{Tighe}},
  \bibinfo{journal}{Soft Matter} \textbf{\bibinfo{volume}{12}},
  \bibinfo{pages}{5450} (\bibinfo{year}{2016}).

\bibitem[{\citenamefont{Dagois-Bohy et~al.}(DOI:
  10.1039/C7SM01846K)\citenamefont{Dagois-Bohy, Somfai, Tighe, and van
  Hecke}}]{dagoisbohy17}
\bibinfo{author}{\bibfnamefont{S.}~\bibnamefont{Dagois-Bohy}},
  \bibinfo{author}{\bibfnamefont{E.}~\bibnamefont{Somfai}},
  \bibinfo{author}{\bibfnamefont{B.~P.} \bibnamefont{Tighe}}, \bibnamefont{and}
  \bibinfo{author}{\bibfnamefont{M.}~\bibnamefont{van Hecke}},
  \bibinfo{journal}{Soft Matter}  (\bibinfo{year}{DOI: 10.1039/C7SM01846K}).

\bibitem[{\citenamefont{Vahabi et~al.}(2017)\citenamefont{Vahabi, Vos,
  de~Cagny, Bonn, Koenderink, and MacKintosh}}]{vahabi17}
\bibinfo{author}{\bibfnamefont{M.}~\bibnamefont{Vahabi}},
  \bibinfo{author}{\bibfnamefont{B.~E.} \bibnamefont{Vos}},
  \bibinfo{author}{\bibfnamefont{H.~C.} \bibnamefont{de~Cagny}},
  \bibinfo{author}{\bibfnamefont{D.}~\bibnamefont{Bonn}},
  \bibinfo{author}{\bibfnamefont{G.~H.} \bibnamefont{Koenderink}},
  \bibnamefont{and}
  \bibinfo{author}{\bibfnamefont{F.}~\bibnamefont{MacKintosh}},
  \bibinfo{journal}{arXiv preprint arXiv:1712.02733}  (\bibinfo{year}{2017}).

\bibitem[{\citenamefont{\AA{}str\"om et~al.}(2000)\citenamefont{\AA{}str\"om,
  M\"akinen, Alava, and Timonen}}]{astrom00}
\bibinfo{author}{\bibfnamefont{J.~A.} \bibnamefont{\AA{}str\"om}},
  \bibinfo{author}{\bibfnamefont{J.~P.} \bibnamefont{M\"akinen}},
  \bibinfo{author}{\bibfnamefont{M.~J.} \bibnamefont{Alava}}, \bibnamefont{and}
  \bibinfo{author}{\bibfnamefont{J.}~\bibnamefont{Timonen}},
  \bibinfo{journal}{Phys. Rev. E} \textbf{\bibinfo{volume}{61}},
  \bibinfo{pages}{5550} (\bibinfo{year}{2000}).

\bibitem[{\citenamefont{Wilhelm and Frey}(2003)}]{wilhelm03}
\bibinfo{author}{\bibfnamefont{J.}~\bibnamefont{Wilhelm}} \bibnamefont{and}
  \bibinfo{author}{\bibfnamefont{E.}~\bibnamefont{Frey}},
  \bibinfo{journal}{Phys. Rev. Lett.} \textbf{\bibinfo{volume}{91}},
  \bibinfo{pages}{108103} (\bibinfo{year}{2003}).

\bibitem[{\citenamefont{Head et~al.}(2003)\citenamefont{Head, Levine, and
  MacKintosh}}]{head03}
\bibinfo{author}{\bibfnamefont{D.~A.} \bibnamefont{Head}},
  \bibinfo{author}{\bibfnamefont{A.~J.} \bibnamefont{Levine}},
  \bibnamefont{and}
  \bibinfo{author}{\bibfnamefont{F.}~\bibnamefont{MacKintosh}},
  \bibinfo{journal}{Physical review letters} \textbf{\bibinfo{volume}{91}},
  \bibinfo{pages}{108102} (\bibinfo{year}{2003}).

\bibitem[{\citenamefont{Das et~al.}(2012)\citenamefont{Das, Quint, and
  Schwarz}}]{das12}
\bibinfo{author}{\bibfnamefont{M.}~\bibnamefont{Das}},
  \bibinfo{author}{\bibfnamefont{D.}~\bibnamefont{Quint}}, \bibnamefont{and}
  \bibinfo{author}{\bibfnamefont{J.}~\bibnamefont{Schwarz}},
  \bibinfo{journal}{PloS One} \textbf{\bibinfo{volume}{7}},
  \bibinfo{pages}{e35939} (\bibinfo{year}{2012}).

\bibitem[{\citenamefont{Goodrich et~al.}(2015)\citenamefont{Goodrich, Liu, and
  Nagel}}]{goodrich15}
\bibinfo{author}{\bibfnamefont{C.~P.} \bibnamefont{Goodrich}},
  \bibinfo{author}{\bibfnamefont{A.~J.} \bibnamefont{Liu}}, \bibnamefont{and}
  \bibinfo{author}{\bibfnamefont{S.~R.} \bibnamefont{Nagel}},
  \bibinfo{journal}{Phys. Rev. Lett.} \textbf{\bibinfo{volume}{114}},
  \bibinfo{pages}{225501} (\bibinfo{year}{2015}).

\bibitem[{\citenamefont{Reid et~al.}(2018)\citenamefont{Reid, Pashine, Wozniak,
  Jaeger, Liu, Nagel, and de~Pablo}}]{reid18}
\bibinfo{author}{\bibfnamefont{D.~R.} \bibnamefont{Reid}},
  \bibinfo{author}{\bibfnamefont{N.}~\bibnamefont{Pashine}},
  \bibinfo{author}{\bibfnamefont{J.~M.} \bibnamefont{Wozniak}},
  \bibinfo{author}{\bibfnamefont{H.~M.} \bibnamefont{Jaeger}},
  \bibinfo{author}{\bibfnamefont{A.~J.} \bibnamefont{Liu}},
  \bibinfo{author}{\bibfnamefont{S.~R.} \bibnamefont{Nagel}}, \bibnamefont{and}
  \bibinfo{author}{\bibfnamefont{J.~J.} \bibnamefont{de~Pablo}},
  \bibinfo{journal}{Proceedings of the National Academy of Sciences} p.
  \bibinfo{pages}{201717442} (\bibinfo{year}{2018}).

\end{thebibliography}

\end{document}